\documentclass[%
 aps,
 jmp,%
 amsmath,amssymb,
 reprint,runinaddress,
 showkeys,floatfix, twocolumn
]{revtex4-1}
\usepackage{graphicx}
\usepackage{dcolumn}
\usepackage{bm}
\usepackage{siunitx}
\usepackage{color}
\usepackage[english]{babel}
\newcommand{\bk}{\mathbf{k}}
\newcommand{\bq}{\mathbf{q}}

\newcommand{\ket}[1]{\big|#1\big>}

\usepackage{letltxmacro}

\LetLtxMacro{\ORIGselectlanguage}{\selectlanguage}
\makeatletter
\DeclareRobustCommand{\selectlanguage}[1]{%
  \@ifundefined{alias@\string#1}
    {\ORIGselectlanguage{#1}}
    {\begingroup\edef\x{\endgroup
       \noexpand\ORIGselectlanguage{\@nameuse{alias@#1}}}\x}%
}
\newcommand{\definelanguagealias}[2]{%
  \@namedef{alias@#1}{#2}%
}
\makeatother
\definelanguagealias{en}{english}
\definelanguagealias{En}{english}
\definelanguagealias{EN}{english}
\definelanguagealias{{EN}}{english}

\begin{document}
\preprint{AIP/123-QED}
\title{Prospects and limitations of transition-metal dichalcogenide laser gain materials
}

\author{F. Lohof}
\affiliation{Institute for Theoretical Physics, Otto-Hahn-Allee 1, P.O. Box 330440, University of Bremen}
\author{A. Steinhoff}
\affiliation{Institute for Theoretical Physics, Otto-Hahn-Allee 1, P.O. Box 330440, University of Bremen}
\author{M. Lorke}
\affiliation{Institute for Theoretical Physics, Otto-Hahn-Allee 1, P.O. Box 330440, University of Bremen}
\author{M. Florian}
\affiliation{Institute for Theoretical Physics, Otto-Hahn-Allee 1, P.O. Box 330440, University of Bremen}
\author{D. Erben}
\affiliation{Institute for Theoretical Physics, Otto-Hahn-Allee 1, P.O. Box 330440, University of Bremen}
\author{F. Jahnke}
\affiliation{Institute for Theoretical Physics, Otto-Hahn-Allee 1, P.O. Box 330440, University of Bremen}
\author{C. Gies}
\affiliation{Institute for Theoretical Physics, Otto-Hahn-Allee 1, P.O. Box 330440, University of Bremen}

\date{\today}
\begin{abstract}
Nanolasers operate with a minimal amount of active material and low losses. In this regime, single layers of transition-metal dichalcogenides (TMDs) are being investigated as next generation gain materials due to their high quantum efficiency.   
We provide results from microscopic gain calculations of highly excited TMD monolayers and specify requirements to achieve lasing with four commonly used TMD semiconductors. Our approach includes band-structure renormalizations due to excited carriers that trigger a direct-to-indirect band-gap transition. As a consequence, we predict a rollover for the gain that limits the excitation regime where laser operation is possible. A parametrization of the peak gain is provided that is used in combination with a rate-equation theory to discuss consequences for experimentally accessible laser characteristics.   
\end{abstract}
\pacs{Valid PACS appear here}
\maketitle
\section{Introduction}
Nanolasers receive much attention due to their promise of energy efficient light generation and integrability in optoelectronic devices \cite{noda_seeking_2006}. Resonators, such as microdisc or photonic-crystal cavities \cite{tamboli_room-temperature_2007, athanasiou_room_2014,altug_ultrafast_2006}, confine light to a volume of its cubic wavelength. Within this mode volume the light-matter interaction is strongly increased, such that the emission into the laser mode is favored over loss channels, giving rise to the high efficiency that is sought after in nanolasers \cite{ellis_ultralow-threshold_2011,chow_emission_2014}.
Gain materials from atomically thin semiconductors have moved into the focus of investigations for their potential in laser applications \cite{wang_electronics_2012}. Monolayer transition-metal dichalcogenides (TMDs) posses a direct band gap at the K point and exceptionally strong Coulomb effects even at ambient conditions, causing the excitonic states to lie in the optical to infrared range about 0.5\,eV below the single-particle band gap \cite{ugeda_giant_2014,wang_electronics_2012}. In different combinations of material and cavity-realizations, first demonstrations of TMD nanolasers have appeared in the literature both at cryogenic and room temperature, using WS$_2$, WSe$_2$, and MoTe$_2$ as active materials \cite{salehzadeh_optically_2015, wu_monolayer_2015, ye_monolayer_2015, li_room-temperature_2017}.

In contrast to gain media that are commonly used in nanolasers, such as semiconductor quantum wells \cite{jagsch_quantum_2018, javerzac-galy_excitonic_2018} and quantum dots \cite{strauf_self-tuned_2006, ota_thresholdless_2017}, the band structure and the electronic states of TMD monolayers are highly susceptible to modifications of the dielectric environment. Similar to the effect of a dielectric embedding, the presence of excited carriers causes screening of the Coulomb interaction and band-structure renormalizations, which have been demonstrated to cause a relative shift within the valley structure: With increasing carrier density, the conduction band $\Sigma$ valley between the K and $\Gamma$ points moves below the K point, rendering the semiconductor indirect \cite{steinhoff_efficient_2015, erben_excitation-induced_2018}. A similar transition takes place in the valence band between K and $\Gamma$. It is clear that such behavior is not captured in linear or logarithmic gain models that are typically used to feed rate-equation models to assess TMD laser performance, even though a direct-to-indirect transition has the most severe consequences for laser operation.

Atomically thin gain materials offer new and exciting possibilities in nanolaser design. In contrast to established epitaxially grown semiconductor nanostructures, such as self-organized quantum dots, quantum wires, or quantum wells, they can be positioned deterministically into the evanescent field of photonic-crystal or microdisc cavities. At the same time, sub- and superstrates to the active TMD monolayers have been demonstrated to enable a wide range of tuning possibilities of the electronic and optical properties of the gain medium  \cite{ugeda_giant_2014,latini_excitons_2015,raja_coulomb_2017,steinhoff_exciton_2017}. In particular, pre-structured dielectric and plasmonic environments allow to tailor the local carrier landscape similar to how the photonic landscape can be formed by structuring a photonic crystal \cite{rooney_observing_2017, rosner_two-dimensional_2016, steinke_noninvasive_2017}. Combined advancements in the understanding of material properties and fabrication techniques hold enticing possibilities for integrated photonics applications and light sources.

With the present work we close the gap between currently used phenomenological laser models and excited-state optical properties on a microscopic foundation. The first part of this work introduces gain calculations by solving semiconductor Bloch equations over the whole Brillouin zone using band structure and interaction matrix elements from ab-initio G$_0$W$_0$ calculations. For the driven laser system, we assume that carriers are generated at high densities. Many-particle renormalizations of the electronic properties due to excited carriers cause a TMD-generic rollover of the gain with increasing carrier density. For the commonly used TMDs MoS$_2$, MoSe$_2$, WS$_2$, and WSe$_2$ we provide a parametrization of the material gain that can be used instead of simplified gain models taken from generic quantum-well or bulk semiconductors. In the second part, we introduce a laser rate-equation theory that is suited to describe TMD nanolaser realizations, in which the extended states of the atomically thin semiconductor flakes couple to localized modes of high-Q microcavities, such as photonic-crystal or whispering-gallery mode resonators. In combination with the parametrized gain, these rate equations offer a practical tool for analyzing existing realizations and for guiding future designs of TMD nanolasers. This we demonstrate by discussing the input-output characteristics, gain clamping, and emission quenching due to the aforementioned rollover for MoS$_2$, WS$_2$ and WSe$_2$.

\section{Results and discussion}
\subsection{TMD Material Gain}
\label{sec:TMD gain}

\begin{figure*}[h!]
\begin{center}
\includegraphics[width=1.\textwidth]{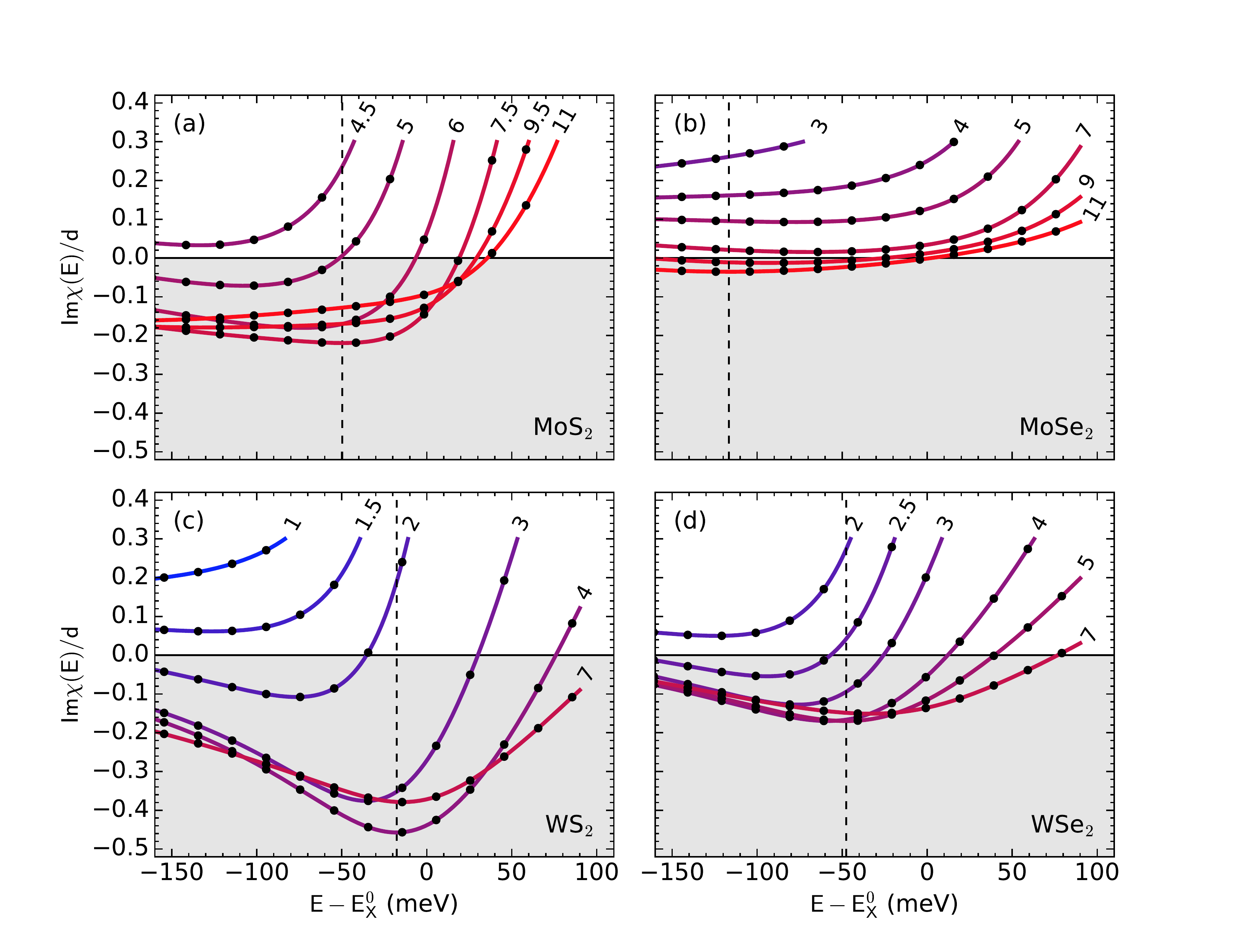}
\caption{Imaginary part of dimensionless optical susceptibility as obtained from the SBE for \textbf{(a)} monolayer MoS$_2$, \textbf{(b)} MoSe$_2$, \textbf{(c)} WS$_2$, and \textbf{(d)} WSe$_2$ on a SiO$_2$ substrate at $T=300$ K and increasing excitation density $N$ given in $10^{13}/$cm$^2$. While a positive value of $\textrm{Im}\,\chi$ implies absorption, optical gain is characterized by a negative imaginary part of the susceptibility (shaded region). The energy axis is chosen relative to the energy $E_{\textrm{X}}^0$ of the A exciton at zero excitation density and low temperatures \cite{cadiz_excitonic_2017}. Vertical lines (dotted) indicate the energies at which maximum gain is achieved for each material.
}
\label{fig:gain}
\end{center}
\end{figure*}
The material gain of an active semiconductor medium is obtained from its optical response to a classical electric field $\textbf{E}(t)$~\cite{chow_semiconductor-laser_1999}, which drives a macroscopic polarization of the medium. This is composed of the microscopic inter-band polarizations $\psi^{\textrm{he}}_{\bk}(t)=\big<a^{\textrm{h}}_{\bk}a^{\textrm{e}}_{\bk}\big>(t)$, with the annihilation operator $a^{\textrm{e/h}}_{\bk}$ of electrons and holes, according to $\textbf{P}(t)=\frac{1}{\mathcal{A}}\sum_{\bk,\textrm{eh}}\psi^{\textrm{he}}_{\bk}(t)(\textbf{d}_{\bk}^{\textrm{eh}})^*+\textrm{c.c.}$ Here, $\mathcal{A}$ is the crystal area and $\textbf{d}^{\textrm{eh}}_{\bk}$ are dipole matrix elements describing the efficiency of light-matter coupling depending on the momentum $\bk$. By taking into account only optical transitions diagonal in momentum space we assume that the optical mode coupling to our gain medium extends over a large number of unit cells such that translational invariance is approximately conserved, which is true even in case of a nanolaser with a mode volume $\sim (\lambda/n)^3$.

In response to a weak optical test field, the linear optical susceptibility is given by
\begin{equation}
\begin{split}
\chi(\omega) = \frac{P(\omega)}{\varepsilon_0 E(\omega)}\,,
\end{split}
\end{equation}
\label{eq:chi}
with $P$ being the polarization in direction of the external field. We assume a circular polarization of the electric field. To obtain the frequency-dependent response of the medium, we use the Fourier transform of the macroscopic polarization $\textbf{P}(t)$, assuming the medium to be in a quasi-equilibrium state.
\par
As the macroscopic polarization stems from the dynamics of electrons and holes, the gain medium has to be described quantum mechanically. On top of that, the inclusion of many-body effects that shape the optical response in the high-excitation regime requires quantum-field theoretical methods. We use the \textit{semiconductor Bloch equations} (SBE), including many-body effects due to excited carriers on GW level. A comprehensive explanation is found in the Appendix A.

The SBE connect the two-particle optical response to single-particle properties such as band structures, electron and hole occupancies and dipole matrix elements. The single-particle band structure, Coulomb, and Dipole matrix elements are obtained on G$_0$W$_0$ level for each TMD material and enter the SBE to capture the multi-valley band structure as well as nonlocal screening effects due to the dielectric environment of the atomically thin layer.
The SBE systematically take into account density-dependent many-body renormalizations of quasi-particle band structures and the corresponding quasi-equilibrium adjustment of electron and hole occupancies. As soon as the combined occupancies of electrons and holes exceed unity for a finite number of momentum states, population inversion is achieved, which is the premise for gain in an electron-hole plasma. In TMD monolayers, efficient light-matter interaction takes place in the K and K$'$ valleys of the band structure \cite{xu_spin_2014}. Hence the amount of optical gain that can be achieved with a certain TMD medium crucially depends on the relative energetic positions of different band-structure valleys that compete with the optically active K and K$'$ valleys for excited carriers. In this respect, there is a fundamental difference between Mo-based and W-based TMD materials since the lowest inter-band transition of the latter is dark due to conduction-band spin-orbit splitting of about $50$ meV \cite{wang_-plane_2017,zhang_magnetic_2017}. Moreover, monolayer WSe$_2$ has been shown to be intrinsically indirect since the $\Sigma$ valley, halfway between K and $\Gamma$, is energetically lower than the K valley \cite{hsu_evidence_2017}. A comparable situation is predicted for MoSe$_2$\cite{erben_excitation-induced_2018}. The `directness' of TMD monolayers is also strongly subject to strain  \cite{conley_bandgap_2013,steinhoff_efficient_2015} and renormalization effects induced by excited carriers \cite{erben_excitation-induced_2018}, which is of pivotal importance for optical gain as discussed in the following.

In Fig.~\ref{fig:gain} we present numerical results for the optical gain in monolayers of four commonly studied TMD semiconductors on top of a SiO$_2$ crystal.
While MoS$_2$ and WS$_2$ exhibit a direct band gap at the K point, MoSe$_2$ and WSe$_2$ are intrinsically indirect and set different preconditions for achieving gain. We focus on optical gain generated by a fully ionized electron-hole plasma by choosing excitation densities above the Mott transition of excitons \cite{steinhoff_exciton_2017, chernikov_population_2015}. For later use we rescale the susceptibility $\chi(\omega)$ with respect to the TMD layer thickness $d=0.6$ nm to obtain a dimensionless quantity.
As a general trend, the material absorption decreases and turns into gain for low to intermediate excitation densities as the population inversion at the direct optical transition in the K  and K$'$ valleys of the quasi-particle band structure builds up. At the same time, the spectral
window where gain is obtained
experiences a blue shift as a net result of several competing effects. While many-particle renormalizations of the band gap yield a red shift of optical transitions \cite{steinhoff_influence_2014, pogna_photo-induced_2016, cunningham_photoinduced_2017}, band filling leads to a blue shift of the Fermi edge, where the maximum population inversion is obtained. The resulting increase of optical band gap, known as Burstein-Moss effect, overcompensates the many-particle renormalization. Besides these common observations, there are important differences between the four TMD materials that define their suitability as gain materials for optoelectronic device applications.
In general, Mo-based TMD materials exhibit larger effective electron and hole masses than W-based materials \cite{kylanpaa_binding_2015}. This leads to broader carrier distributions in $k$-space, which makes it harder to achieve population inversion at the band edge. Consequently, gain sets in at higher carrier densities in Mo-based materials. At the same time, the effective masses of the selenides are larger than those of the sulfides, hampering population inversion in the selenides. This tendency turns even more into a disadvantage due to the intrinsic indirectness of the Se-based TMDs. WS$_2$ can be inverted most easily and has indeed been the first TMD material for which population inversion has been demonstrated in experiment \cite{chernikov_population_2015}.
MoSe$_2$ is ill-suited to provide gain due to the combined detrimental effects of slow hole population buildup and electron drain from the K valley, which is discussed in detail below.
For MoS$_2$ and WSe$_2$, the different band-structure effects due to transition metal and chalcogen atoms compensate to some degree. The onset of gain appears much later in MoS$_2$ due to the large hole mass and an additional drain of holes from the K to the $\Gamma$ valley at elevated carrier densities.

The optical susceptibility shown in Fig.~\ref{fig:gain} can be translated into a hypothetical gain per cm that would be generated by photons propagating laterally through the gain material. The intensity gain per cm $g(\omega)$, which is an important figure of merit for edge emitters, is defined as \cite{chow_semiconductor-laser_1999}:
\begin{equation}
\begin{split}
g(\omega)=-\frac{\omega}{c n_{\textrm{s}} d}\textrm{Im}\,\chi(\omega)
\label{eq:susc_dimless}
\end{split}
\end{equation}
with the refraction index $n_{\textrm{s}}=1.46$ of the SiO$_2$ substrate at optical frequencies. For WS$_2$, we obtain a maximum intensity gain of $3\times 10^4$ cm$^{-1}$, which is about a factor of 4 larger than predicted for GaAs quantum wells  \cite{chow_semiconductor-laser_1999}. This is remarkable given the fact that we deal with an atomically thin layer of material.

A nontrivial effect that is special to atomically thin TMD semiconductors is the tendency to become more indirect semiconductors under carrier excitation, a behavior that is usually associated with the application of strain \cite{erben_excitation-induced_2018,steinhoff_efficient_2015}. This transition impacts the density-dependence of gain at high excitation densities. The gain reaches a maximum at a density $N=7\times10^{13}$ cm$^{-2}$ for MoS$_2$ and $N=4\times10^{13}$ cm$^{-2}$ for WS$_2$ and WSe$_2$ before a drastic rollover is observed. For MoSe$_2$, where almost no gain is obtained, no gain rollover is visible.
\begin{figure}[h]
\begin{center}
\includegraphics[width=0.5\textwidth]{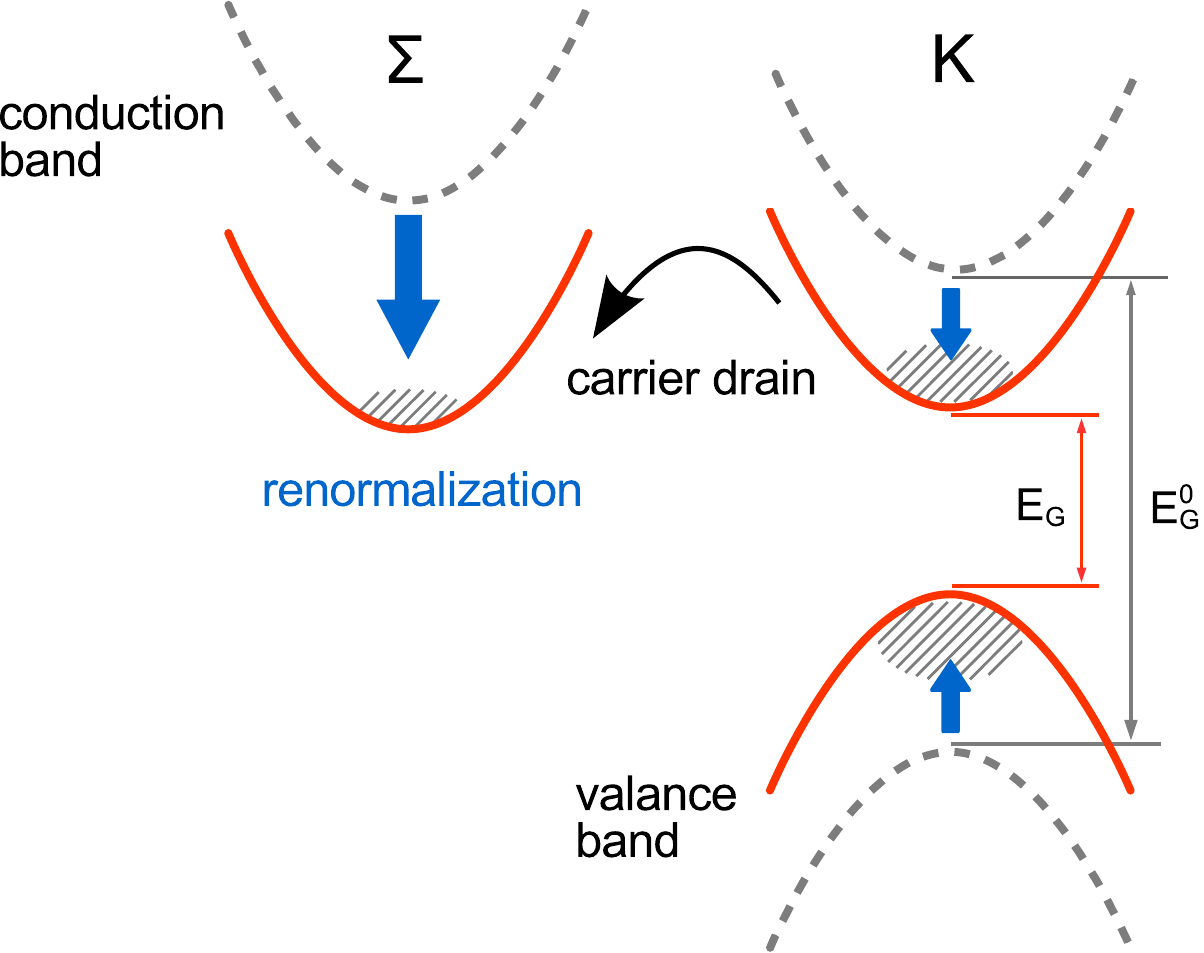}
\caption{Schematic of the band-structure renormalization induced by excited carriers. The direct band gap $E^0_{\textrm{G}}$ at the K point shrinks and assumes the actual value $E_{\textrm{G}}$ at a certain carrier density. At the same time, the conduction-band valley at the $\Sigma$ point is shifted even more strongly, becoming energetically more favorable for electrons than the K point at a critical carrier density. As a consequence, electrons are drained from K to $\Sigma$, thereby being lost to the optically active region around K.}
\label{fig:renorm_scheme}
\end{center}
\end{figure}
%
The mechanism behind the rollover effect is explained in Fig.~\ref{fig:renorm_scheme}. While the general trend under carrier excitation is a redshift of the band structure, efficient electron-hole exchange interaction leads to a blueshift of the K valley relative to the $\Sigma$ valley, halfway between K and $\Gamma$ \cite{erben_excitation-induced_2018}. Thereby the $\Sigma$ point becomes energetically more favorable for electrons than the K point, draining carriers from the optically active region around the K point. This leads to a quenching of population inversion at K above a certain excitation density. In combination with an ever-increasing efficiency of excitation-induced dephasing that leads to a broadening of resonances, the quenching of inversion leads to the observed rollover of the peak gain. As WSe$_2$ and MoSe$_2$ intrinsically exhibit an indirect band gap already on the level of the G$_0$W$_0$-band structures entering the SBE, the carrier drain from $\Sigma$ to K starts at low excitation densities in these materials.
Due to this electron drain to the $\Sigma$ valley, electron populations at the K valley are usually smaller than hole populations. Hence most of the population inversion at elevated carrier densities is provided by holes, so that the unfavorable ordering of conduction-band spins in the W-based materials plays a minor role for our results.

\subsection{TMD Laser Rate Equations}
\label{sec:RE}
Laser rate equations \cite{coldren_diode_1995,bjork_analysis_1991} are an established tool for modeling a wide range of laser devices. Rate equations have also been used in the analysis of TMD monolayer laser devices using either generalized linear \cite{li_room-temperature_2017, ye_monolayer_2015} or logarithmic  \cite{salehzadeh_optically_2015} gain models, or purely excitonic gain  \cite{wu_monolayer_2015}, in which case the rate equations resemble those used for atomic systems. In the following, we provide rate equations suited for nanolasers with TMD monolayer gain interacting with a single, localized cavity mode such as provided by whispering-gallery-mode or photonic-crystal resonators. In particular, instead of relying on a generic gain model, we incorporate the microscopically calculated gain from the previous section. One important reason to do so is to account for effects that arise from the particular electronic properties of the quantum gain materials, such as the direct-to-indirect band-gap transition. Another important reason lies in the definition of the material gain for an extended medium interacting with a localized mode. Conventional rate equations often employ gain models that are based on the concept of propagating modes, resulting in units of 1$/$cm for the gain and the need for introducing a group velocity $v_\mathrm{g}$ in the rate equations \cite{coldren_diode_1995}. The situation is different if the mode function is spatially localized and not propagating \cite{haug_semiconductor_1989, meystre_elements_2007}, which is the case in all of the reported experiments on TMD lasing. Complementary to Eq.~\eqref{eq:susc_dimless}, we define the the resulting material gain $G(N)$ in units of 1$/$ps directly in terms of the dimensionless susceptibility $\chi(\hbar\omega_0,N)$
\begin{equation}\label{eq:gain}
G(N)=-\frac{\omega_0}{d}\mathrm{Im}\left[\chi(\hbar\omega_0,N)\right],
\end{equation}
with $\omega_0$ being the eigenfrequency of the mode. With this, the rate equations for the carrier density $N$ in the TMD monolayer, and the photon density $N_\mathrm{p}$ in the cavity mode are given by
\begin{subequations}
\begin{align}
\dot{N} & = \frac{P}{A_\mathrm{s}} - \frac{V_\mathrm{m}}{A_\mathrm{s}} \Gamma G(N)N_\mathrm{p} \nonumber \\
& \qquad - \frac{V_\mathrm{m}}{A_\mathrm{s}} ( AN+BN^2+CN^3 )  \label{RE1}\\
\dot{N_\mathrm{p}} & = \Gamma G(N)N_\mathrm{p} +\beta BN^2-\frac{N_\mathrm{p}}{\tau_\mathrm{c}}.\label{RE2}
\end{align}\label{RE}
\end{subequations}
Here, $P$ is the carrier generation rate in the area $A_\mathrm{s}$ of active gain material, $\Gamma$ is the optical confinement factor, $V_\mathrm{m}$ is the mode volume and $1/\tau_\mathrm{c}=\omega_0/Q$ is the cavity lifetime with the cavity resonance frequency $\omega_0$ and quality factor $Q$.
$A$, $B$ and $C$ are the coefficients of non-radiative and radiative recombination used in standard rate-equation theory, and $\beta$ is the spontaneous emission coupling factor.

High $\beta$-factors reported in recent publications on TMD lasers are often attributed to increased spontaneous emission into the lasing mode due to a large Purcell factor. However, Purcell's original expression (proportional to $Q/V_\mathrm{m}$) is derived under the assumption of $\Delta\omega_\mathrm{e}\ll\Delta\omega_\mathrm{c}$ with $\Delta\omega_\mathrm{e}$ and $\Delta\omega_\mathrm{c}$ being the spectral linewidths of emitter and cavity mode, respectively \cite{romeira_purcell_2018, baba_photonic_1997}. This condition can indeed be satisfied e.g.~for atomic systems or quantum dots in optical cavities. Here we consider room-temperature setups of extended gain media in high-$Q$ cavities in which the situation is typically reversed, i.e.~$\Delta\omega_\mathrm{e}\geq\Delta\omega_\mathrm{c}$ and instead of a Purcell-effect a high $\beta$ is achieved by suppression of leaking modes and large mode separation.

Care must be taken in ensuring consistency of dimension: The carrier density $N$ is an area density in reference to the active two-dimensional TMD gain material. The photon density is defined with respect to the mode volume $V_\mathrm{m}$.
The mismatch in dimension between the gain material and the mode extension requires to introduce the factor $V_\mathrm{m}/A_\mathrm{s}$ \cite{coldren_diode_1995}. A consequence of this reconciliation is merely a rescaling of the excatiation power axis ($P\propto V_\mathrm{m}$ for stationary solutions), while the relative gain and loss contributions in the rate equations remain unchanged. While the mode volume is well defined, it appears sensible to identify the area $A_\mathrm{s}$ with the area in which carriers are excited within the TMD flake. Physically, it is therefore defined by a relation between the spot diameter of the exciting laser, and the carrier diffusion length.

The optical confinement factor measures the spatial overlap of the gain medium with the cavity mode. It is given by
\begin{equation}
\Gamma=\frac{\int_\mathrm{Layer}\epsilon(\bm{r})|E_{||}(\bm{r})|^2\mathrm{d}r}{\int_V\epsilon(\bm{r})|E(\bm{r})|^2\mathrm{d}r},\label{Confinement_factor}
\end{equation}
where $E_{||}(\bm{r})$ is the in-plane electric field of the cold cavity \cite{haug_semiconductor_1989, huang_analysis_1996}, and the denominator is proportional to the cavity mode volume $V_m$. Together with the material gain it constitutes the modal gain $\Gamma G(N)$ which enter the rate equations.
Neglecting the spontaneous-emission contribution in Eq.~\eqref{RE2} one easily sees that the threshold condition of gain compensating the cavity losses depends on the product $\Gamma Q$. This gives the optical confinement factor a particular relevance, as it must be determined in order to uniquely specify $Q$.


\label{sec:disc}
 \begin{figure}[h]
 \begin{center}
 \includegraphics[width=0.5\textwidth]{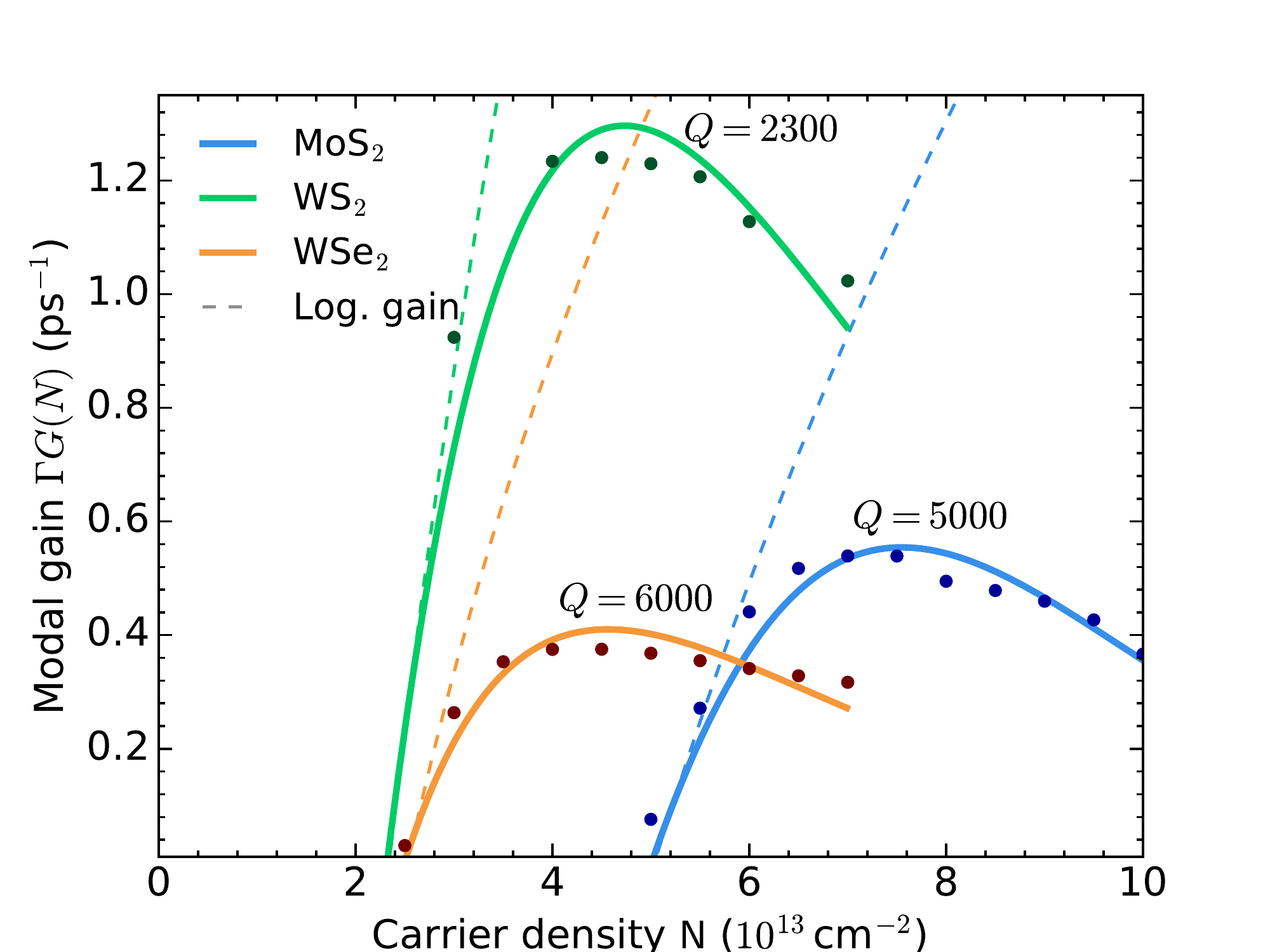}
 \caption{Modal gain $\Gamma G(N)$ as a function of carrier density for MoS$_2$, WS$_2$ and WSe$_2$ at energies where maximum gain is achieved. All materials show a characteristic rollover limiting gain from the single layer.  In WS$_2$ gain sets in at lowest carrier densities and takes on the greatest value out of all four TMD materials. While MoS$_2$ requires much larger excitation densities to achieve gain, it still exhibits a larger peak gain than WSe$_2$. Due to the minuscule amount of gain obtained from MoSe$_2$ it has been omitted from this figure. Solid lines indicate the parametrization according to Eq.~\eqref{gain_fit}. Dashed lines show logarithmic fits to the gain curves at low densities. For the used parameters, the minimal $Q$-factors necessary to achieve lasing are listed for each material.}
 \label{fig:gain_at_energy}
 \end{center}
 \end{figure}

\subsection{Input-output characteristics}
\label{sec:lasercharacteristics}

We evaluate the rate-equation model using the microscopically calculated gain as input via Eq.~\eqref{eq:gain}. We begin by analyzing the modal gain $\Gamma G(N)$ of MoS$_2$, WS$_2$ and WSe$_2$ as a function of carrier density. Fig.~\ref{fig:gain_at_energy} is obtained by collecting the results in Fig.~\ref{fig:gain} at fixed energies (vertical lines) chosen to maximize the gain for each TMD material. Gain rollover is clearly visible at higher carrier densities, the origin of which has been discussed in the previous section, and which puts an intrinsic upper limit on the amount of gain that is achievable from a single TMD monolayer.

The behavior of the microscopically calculated gain is well reproduced by a fit formula
\begin{multline}
G(N)=\\ \omega_0\left(a(N-N_0)e^{-2b(N+N_1)(N-N_0)-b(N-N_0)^2}\right) \label{gain_fit}
\end{multline}
that combines a linear gain, characterized by the differential gain at low densities $a$ and the transparency density $N_0$, with a shifted Gaussian to account for the rollover characteristic predicted by the full calculations. Often linear or logarithmic gain models serve in rate-equation theories and have also been used for TMD gain materials. Logarithmic fits are shown as dashed lines in Fig.~\ref{fig:gain_at_energy}, indicating reasonable agreement in the low-density regime but significant deviations and the absence of the rollover behavior at elevated carrier densities.  More details and the parameter values to be used in Eq.~\eqref{gain_fit} are given in the Appendix B.

\begin{figure*}[h]
 \begin{center}
 \includegraphics[width=1.0\textwidth]{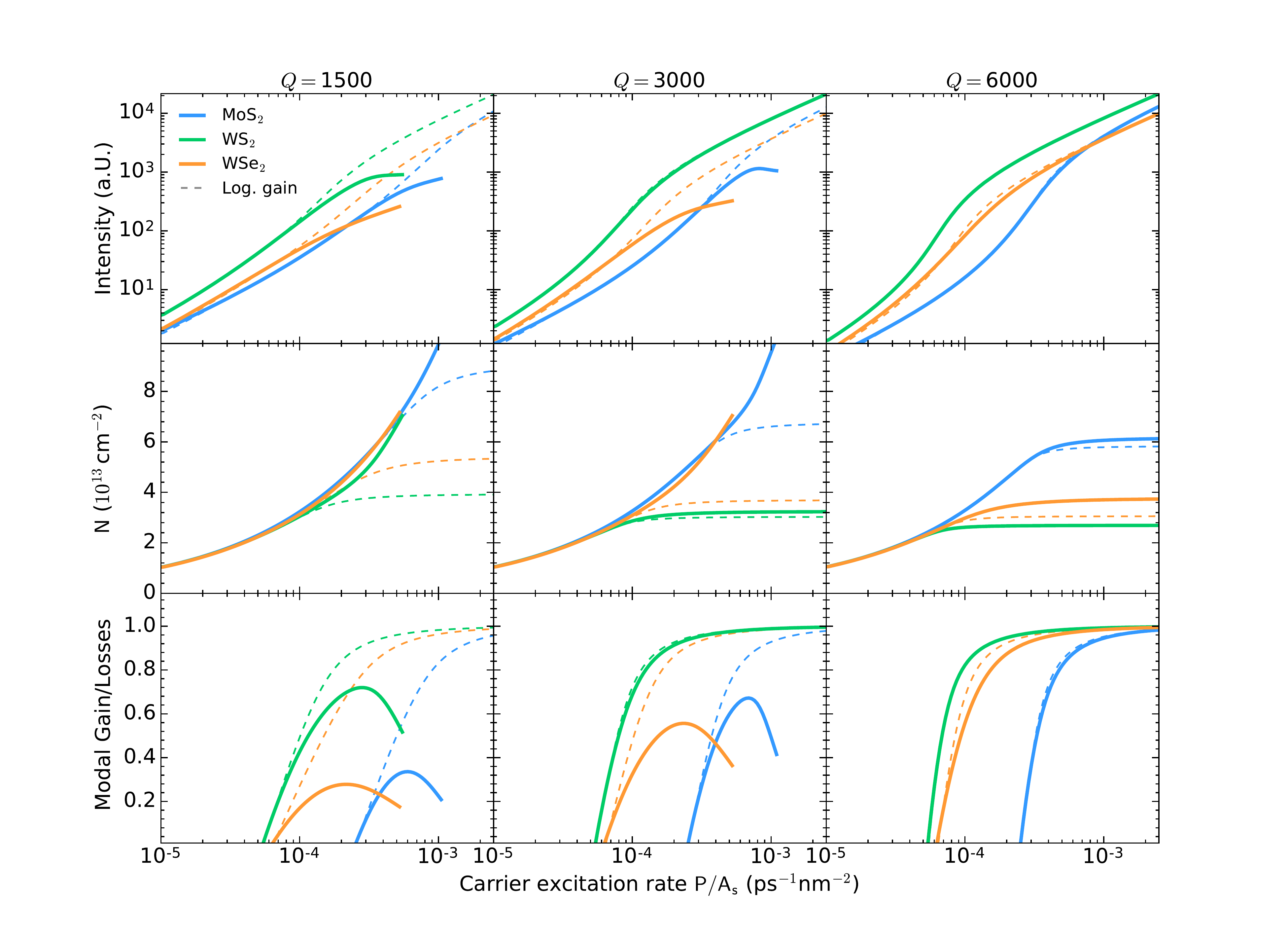}
 \caption{Input-output curves (top), carrier densities (middle) and the ratio  of modal gain and cavity losses (bottom) for increasing $Q$-factors of 1500, 3000 and 6000. Results are obtained from Eqs.~\eqref{RE} using input from the material realistic gain calculation as described in Sec.~\ref{sec:TMD gain} for MoS$_2$, WS$_2$ and WSe$_2$. Dashed lines use a the logarithmic gain fit indicated in Fig.~\ref{fig:gain_at_energy}. In WS$_2$ the lasing threshold is reached at lowest excitation rate and the lowest $Q$-factor is required as the maximum gain is larger in comparison to MoS$_2$ and WSe$_2$. If the product $\Gamma Q$ falls below a material specific value, all materials experience a gain rollover (left panels) and lasing cannot be achieved by stronger pumping. With a logarithmic gain model the threshold can be reached at sufficient pumping and the material specific lasing signatures are no longer reflected in the calculations. Parameters are chosen close to those in \cite{ye_monolayer_2015}.}
 \label{fig:IO_curves}
 \end{center}
 \end{figure*}

Laser rate equations lend themselves to countless possibilities of parameter studies. When used with good intuition, they conversely allow to fit experimental data in order to extract parameters for device characterization. Here, we pursue the objective of illustrating how the material gain plays out in the rate equations to shape the input-output characteristics. We choose typical parameters based  on situations found in the recent experiments (see Table~\ref{tab:paras}). In particular, an optical confinement factor $\Gamma=9 \times 10^{-4}$ is used, which is in between values that have been reported for photonic-crystal and microdisc cavities \cite{li_room-temperature_2017,ye_monolayer_2015}. We refrain from adding Shockley-Read-Hall and Auger losses, as they act on top of the rollover that is intrinsic to the material gain. When the equations are tailored to match a specific experiment, appropriate values for $\Gamma$ and the $ABC$ coefficients must be determined ideally from independent calculations and measurements.
\begin{table}[h!]
\centering
\begin{tabular}{ll}
    \hline\noalign{\smallskip}
 $\Gamma$ & $9\times 10^{-4}$ \\
 $\beta$ & 0.1 \\
 $A$ & 0 \\
 $B$ & $6\times 10^{-6}\,\mathrm{nm}\,\mathrm{ps}^{-1}$ \\
 $C$ & 0 \\
 $V_\mathrm{m}$ & $1.4\times 10^8\,\mathrm{nm}^3$ \\
 $A_\mathrm{s}$ & $5.6\times 10^6\,\mathrm{nm}^2$ \\
\hline
\end{tabular}
\caption{Parameters used for the results in Fig.~\ref{fig:IO_curves}. Note that while the quantities $V_\mathrm{m}$ and $A_\mathrm{s}$ possess physical meaning \cite{ye_monolayer_2015}, they merely define a scale for the excitation-rate axis while leaving the input-output characteristics unchanged.}
\label{tab:paras}
\end{table}

The rate equations \eqref{RE} are solved for their stationary values using a continuous pump. Results for MoS$_2$ (blue) WS$_2$ (green) and WSe$_2$ (orange) are shown in Fig.~\ref{fig:IO_curves} for three different cavity $Q$ factors. At first glance, all input-output curves (top panels) exhibit a non-linearity comparable to what has been observed in Refs \citenum{li_room-temperature_2017, wu_monolayer_2015, ye_monolayer_2015}.  While the non-linearity may hint towards laser operation at higher pump rates, the absence or presence of clamping of the carrier densities (middle panels) reveals that lasing is only achieved in some of the cases: At $Q=1500$ population clamping is absent in all materials and indicates operation below threshold. For the chosen parameters, the gain obtainable from a single WS$_2$ monolayer requires a minimal cavity-$Q$ of about 2400. For a $Q$ factor of 3000, population clamping is still absent in MoS$_2$ and WSe$_2$. Due to their lower peak gain, lasing is only achieved at a higher $Q$ factors of 5000 and 6000 for MoS$_2$ and WSe$_2$, respectively. The top right panel also shows that the laser threshold in WS$_2$ and WSe$_2$ sets in at significantly lower threshold pump rates than for MoS$_2$. The reason lies in the aforementioned lower effective hole mass in those materials, causing a more localized carrier population aiding the formation of inversion. Therefore, WS$_2$ and WSe$_2$ can play out their advantages towards more energy-efficient device operation if the quality factor of the cavity permits.

As discussed above, it is the product of confinement factor and quality factor $\Gamma Q$ that must exceed a material-specific value for gain to compensate the cavity losses and initialize lasing. Specifically, the necessary values for $\Gamma Q$ are  2.18 for WS$_2$, 4.59 for MoS$_2$, and 5.88 for WSe$_2$. The overall behavior is also reflected in the bottom panels of Fig.~\ref{fig:IO_curves}, where the ratio of modal gain and cavity losses is shown. At a value of 1 gain exactly compensates the losses. For low $Q$, i.e.~insufficent $\Gamma Q$, the ratio mirrors the characteristic rollover observed in Fig.~\ref{fig:gain_at_energy} and remains below unity. Note that with a logarithmic gain model (dashed lines) lasing can be achieved for all $Q$ under sufficient pumping due to the absence any mechanism limiting the maximum gain. If the modal gain is sufficiently high to facilitate stimulated emission, gain clamping is observed and the deviations between the material-realistic gain model and the logarithmic fit become small (right panel in Fig.~\ref{fig:IO_curves}). In addition to using high-$Q$ cavities, optimizing the confinement factor $\Gamma$ is a viable way to fulfill the threshold condition.



\section{Conclusion}
Monolayer TMDs are currently being investigated for their potential as ultra-thin gain materials in nanolasers. We have performed gain calculations for four of the commonly used TMD semiconductors on a material-realistic footing. As a consequence of strong band-structure renormalizations and due to excitation-induced dephasing, the TMD material gain exhibits a rollover with increasing carrier density that cannot be captured by linear or logarithmic gain models. To remedy this limitation, we provide a parametrization of the microscopically calculated gain to replace these simple approaches.
We have introduced and evaluated a rate-equation theory for the interaction of an extended two-dimensional gain material with a localized mode. For current device specifications, a single MoS$_2$, WS$_2$ or WSe$_2$ layer provides sufficient gain to unambiguously cross the laser threshold at cavity-$Q$ factors of about 2400-6000. WS$_2$ turns out to provide both the lowest threshold density and the largest maximum gain. Extrapolating results we have obtained for the sulfides and selenides, we expect transition-metal ditellurides to provide only a small plasma gain due to their large effective masses.

An equally important role as the $Q$-factor is given by the confinement factor that defines the spatial overlap of the TMD flake with the localized mode. Cavity geometries that allow an embedding of the TMD flake rather than its weak coupling to the evanescent field exhibit larger confinement factors and can, therefore, operate at lower $Q$ factors. As we have shown, while gain clamping is a strong indicator for laser operation, care must be taken when lasing operation is inferred from the input-output characteristics alone. Photon correlation measurements are still pending for TMD-based nanolasers and will finally help to provide conclusive evidence for laser operation \citep{chow_emission_2014}. As a prospect for future nanolaser realizations with TMD monolayer as active gain material, the achievable gain could be greatly enhanced by stacking several TMD monolayers separated e.g.~by layers of hexagonal boron-nitride. Due to the atomic dimension of the layer thickness, the light-matter coupling is expected to change little from layer to layer, so that vertical heterostructures promise a significant advantage over single monolayer gain materials.

It is known that the Mott density, above which excitons are fully ionized into unbound electrons and holes, is lowered by screening of the Coulomb interaction due to a dielectric environment of the TMD monolayer \cite{steinhoff_exciton_2017}.
One might be tempted to use this fact to lower the threshold current of the nanolaser by encapsulating the active monolayer into material with a large dielectric constant. However, the Mott transition is a necessary but no sufficient condition for population inversion. The carrier density at which inversion is achieved mainly depends on characteristics of the quasi-particle band structure, which is not very susceptible to the dielectric environment but rather to strain applied to the TMD monolayer.

\section*{Acknowledgement}
The authors thank Tim Wehling and Malte R\"osner for providing ab-initio band-structure calculations and Coulomb matrix elements that have been used as input for the gain calculations. Funding from the DFG (``Deutsche Forschungsgemeinschaft'') via the graduate school ``Quantum-Mechanical Material Modeling'' is gratefully acknowledged.
 
  
 \appendix
\section{Semiconductor Bloch Equations}
Applying the technique of nonequilibrium Green functions, the equation of motion for microscopic inter-band polarizations known as \textit{semiconductor Bloch equations} (SBE) can be derived \cite{jahnke_linear_1997, schafer_semiconductor_2002, manzke_quantum_2003}. As shown in detail in Ref.~\citenum{erben_excitation-induced_2018}, the SBE are transformed into frequency space taking into account many-body effects due to excited carriers on a GW-level:
\begin{equation}
\begin{split}
 \Big(\hbar\omega-\varepsilon^{\textrm{HF,h}}_{\bk}-&\varepsilon^{\textrm{HF,e}}_{\bk} -\Delta^{\textrm{eh}}_{\bk}(\omega)+i\gamma^{\textrm{El-Ph}}\Big)\psi_{\bk}^{\textrm{he}}(\omega) \\
 +\Big(1-f_{\bk}^{\textrm{KMS,e}}-&f_{\bk}^{\textrm{KMS,h}}\Big)\Big(\mathbf{d}_{\bk}^{eh}\cdot \mathbf{E}(\omega)
 +\frac{1}{\mathcal{A}}\sum_{\bk'}V_{\bk\bk'\bk\bk'}^{\textrm{ehhe}}\psi_{\bk'}^{\textrm{he}}(\omega)\Big) \\
 &+\frac{1}{\mathcal{A}}\sum_{\bk'}V_{\bk\bk'}^{\textrm{eff,eh}}(\omega)\psi_{\bk'}^{\textrm{he}}(\omega) =0\,.
\label{eq:SBE_GW_freq}
\end{split}
\end{equation}
Just as the well-known Bethe-Salpeter equation in screened ladder approximation  \cite{strinati_effects_1984,bornath_two-particle_1999,kremp_quantum_2005}, the SBE on GW-level describe two-particle states in the presence of a dynamically screened carrier-carrier interaction. The single-particle energies $\varepsilon^{\textrm{HF,e/h}}_{\bk}$ contain renormalization effects on a Hartree-Fock level including the exchange interaction between electrons and holes  \cite{erben_excitation-induced_2018}. The Pauli-blocking term in the second line of Eq.~(\ref{eq:SBE_GW_freq}) defined by the electron and hole occupancies $f_{\bk}^{\textrm{KMS,e/h}}$ is responsible for population inversion of a certain momentum state $\bk$. The occupancies are given in quasi-equilibrium by Kubo-Martin-Schwinger (KMS) distribution functions \cite{kremp_quantum_2005} that emerge as Fermi distribution functions weighted by quasi-particle spectral functions. The second line also contains the light-matter coupling term $\textbf{d}\cdot\textbf{E}$ of the materials' dipoles to the external field as well as a two-body interaction term facilitating excitonic resonances in the optical response. As we focus on the microscopic description of effects induced by excited carriers, we include dephasing contributions due to carrier-phonon interaction on a phenomenological level by adding a constant imaginary part $\gamma^{\textrm{El-Ph}}$ to the quasi-particle energies. The constant is chosen as $\gamma^{\textrm{El-Ph}}=20$ meV.
All many-body effects induced by carrier-carrier interaction beyond the Hartree-Fock level are contained in the correlation terms
\begin{equation}
\begin{split}
V_{\bk\bk'}^{\textrm{eff,eh}}(\omega)&=i\hbar\int_{-\infty}^{\infty}\frac{d\omega'}{2\pi}\Bigg\{\\
    &\frac{(1-f^{\textrm{h}}_{\bk}+n_{\textrm{B}}(\omega'))2i \textrm{Im}\,W^{\textrm{ret,ehhe}}_{\bk\bk'\bk\bk'}(\omega')}{\hbar\omega-\varepsilon^{\textrm{h}}_{\bk}-\varepsilon^{\textrm{e}}_{\bk'}+i\Gamma^{\textrm{h}}_{\bk}+i\Gamma^{\textrm{e}}_{\bk'}-\hbar\omega'} \\
   + &\frac{(1-f^{\textrm{e}}_{\bk}+n_{\textrm{B}}(\omega'))2i \textrm{Im}\,W^{\textrm{ret,ehhe}}_{\bk\bk'\bk\bk'}(\omega')}{\hbar\omega-\varepsilon^{\textrm{e}}_{\bk}-\varepsilon^{\textrm{h}}_{\bk'}+i\Gamma^{\textrm{e}}_{\bk}+i\Gamma^{\textrm{h}}_{\bk'}-\hbar\omega'}\Bigg\}\,,
\label{eq:Veff}
\end{split}
\end{equation}
with the Bose distribution function $n_{\textrm{B}}(\omega)$ and Fermi distribution function $f^{\textrm{e/h}}_{\bk}$, and
\begin{equation}
\begin{split}
\Delta^{\textrm{eh}}_{\bk}(\omega)&=\Sigma_{\bk}^{\textrm{MW},\textrm{ret,e}}(\hbar\omega-\varepsilon^{\textrm{h}}_{\bk}+i\Gamma^{\textrm{h}}_{\bk})\\&+\Sigma_{\bk}^{\textrm{MW},\textrm{ret,h}}(\hbar\omega-\varepsilon^{\textrm{e}}_{\bk}+i\Gamma^{\textrm{e}}_{\bk}),
\label{eq:Delta}
\end{split}
\end{equation}
where the Montroll-Ward (MW) self-energy 
\begin{equation}
\begin{split} 
  & \Sigma_{\bk}^{\textrm{MW},\textrm{ret},\lambda}(\omega) = i\hbar\int_{-\infty}^{\infty}\frac{d\omega'}{2\pi}\\
            \frac{1}{\mathcal{A}}\sum_{\bk'}&\frac{(1-f^{\lambda}(\omega-\omega')+n_{\textrm{B}}(\omega'))2i\,\textrm{Im}\,W^{\textrm{ret},\lambda\lambda\lambda\lambda}_{\bk\bk'\bk\bk'}(\omega') }{\hbar\omega-\varepsilon^{\lambda}_{\bk'}+i\Gamma^{\lambda}_{\bk'}-\hbar\omega'}
    \label{eq:MW}
\end{split}
\end{equation}
is responsible for quasi-particle renormalizations according to
\begin{equation}
\varepsilon^{\lambda}_{\bk} = \varepsilon_{\bk}^{0,\lambda}+\Sigma_{\bk}^{\textrm{HF},\lambda}+\textrm{Re}\, \Sigma_{\bk}^{\textrm{MW},\textrm{ret},\lambda}(\omega)\Big|_{\omega=\varepsilon^{\lambda}_{\bk}/\hbar} \,.
\label{eq:quasip}
\end{equation}
The corresponding quasi-particle broadening follows from the imaginary part of the MW self-energy:
\begin{equation}
\begin{split}
&\Gamma^{\lambda}_{\bk} = -\textrm{Im}\, \Sigma_{\bk}^{\textrm{MW},\textrm{ret},\lambda}(\omega)\Big|_{\omega=\varepsilon^{\lambda}_{\bk}/\hbar} \,.
\label{eq:quasip_broadening}
\end{split}
\end{equation}
The band-structure energies $\varepsilon_{\bk}^{0,\lambda}$ in the absence of excitation-induced renormalizations are obtained from \textit{ab initio} G$_0$W$_0$ calculations for freestanding TMD monolayers as described first in  Ref.~\citenum{steinhoff_influence_2014}. 
The correlation terms (\ref{eq:Veff})--(\ref{eq:MW}) contain the retarded screened potential $W^{\textrm{ret}}_{\bk\bk'\bk\bk'}=V_{\bk\bk'\bk\bk'}\varepsilon^{-1}_{\textrm{exc},\bk-\bk'}(\omega)$, where the inverse dielectric function $\varepsilon_{\textrm{exc},\bq}^{-1}(\omega)$ describes screening due to excited carriers in random phase approximation (RPA). Additionally, $V_{\bk\bk'\bk\bk'}$ is screened by carriers in filled valence-band states as well as the dielectric environment of the atomically thin layer. Along with G$_0$W$_0$-band structures, Coulomb matrix elements for monolayer TMDs on a substrate are obtained in two steps. First, DFT ground-state data for freestanding monolayers are used to calculate bare Coulomb matrix elements and the RPA dielectric function in a localized basis $\ket{\alpha}$ consisting of Wannier functions with dominant d-orbital character for discrete values of momentum $\bq$. Then, bare matrix elements and the dielectric function including substrate effects are parametrized as function of $|\bq|$ using the \textit{Wannier function continuum electrostatics} (WFCE) approach described in Refs.~\citenum{rosner_wannier_2015,florian_dielectric_2018}. The approach combines a continuum electrostatic model for the screening by the dielectric environment in a heterostructure with a localized description of Coulomb interaction. The actual parametrization is provided in Ref.~\citenum{steinhoff_exciton_2017}. The matrix elements are then transformed from the Wannier into the Bloch basis. As discussed in Ref.~\citenum{florian_dielectric_2018}, TMD layers do not form perfect interfaces with surrounding dielectric layers in a van der Waals heterostructure but are separated by inter-layer gaps on the sub-nm scale. We assume here that the TMD monolayers are placed on a SiO$_2$ crystal with dielectric constant $\varepsilon=3.9$ and an inter-layer gap of $0.3\,$nm, which has been found to be a reasonable value, see Ref.~\citenum{florian_dielectric_2018} and the references therein. Besides the screening of Coulomb interaction between excited carriers inside the TMD layer, dielectric screening due to the environment modifies the TMD band structure. We calculate the correction of the freestanding G$_0$W$_0$-band structure using a static GW self-energy with an interaction screened by the difference of a dielectric function with and without the environment  \cite{florian_dielectric_2018}. The SBE (\ref{eq:SBE_GW_freq}) are solved numerically by matrix inversion for each frequency to obtain the frequency-dependent microscopic polarization and, thereby, the optical susceptibility using Eq.~(1) in the main document.
\nocite{*}

\section{Gain parameterization}
With Eq.~(6) in the main text, we have given a fit function for the peak gain as a function of carrier density. The purpose of this parameterization is to provide simple means to use the results of the microscopic gain TMD calculations in rate-equation approaches, rather than relying on linear or logarithmic gain models that fail to capture the material properties especially at elevated carrier densities. A particular problem of using such simpler gain models in rate equations is that lasing can always be achieved if the pumping is sufficiently strong, as the only process limiting the increase of gain with increasing carrier density are Auger losses via the $C$ coefficient.

Eq.~\eqref{gain_fit} is based on a linear fit model, where $a$ and $N_0$ correspond to the differential gain at low densities and the transparency density. The linear gain is modified by a shifted Gaussian to account for the gain rollover that we have identified as a characteristic for the TMD monolayer materials. The mode eigenfrequencies enter the material gain together with further parameters that are given in Table \ref{tab:fit_paras}. The carrier density $N$ in Eq.~\eqref{gain_fit} has units $1/\mathrm{nm}^2$ while the parameters $a$, $N_0$, $b$ and $N_1$ are assumed to have units consistent with $G(N)$ having units of $1/\mathrm{ps}$. The gain fit is valid for the density ranges $N\in [0.1, 1.1 ]\, \mathrm{nm}^{-2}$ (MoS$_2$) and $N\in [0.1, 0.7 ]\, \mathrm{nm}^{-2}$ (WS$_2$ and WSe$_2$).

\vspace{1cm}

\begin{table}[h!]
\caption{Parameters for the gain fit in Eq.~\eqref{gain_fit}.}
\centering
\begin{tabular}{lccccccc} 
\hline
\noalign{\smallskip}
 & $\omega_0\,\mathrm{(ps}^{-1}\mathrm{)}$ & $a$ & $N_0$ & $b$ & $N_1$ \\  
\cline{2-6}
\noalign{\smallskip}
MoS$_2$ & 2749 & 2.025 & 0.5 & 2.6 & 0 \\
WS$_2$ & 3001 & 5.147 & 0.232 & 0.85 & 1.95 \\
WSe$_2$ & 2457 & 2.415 & 0.25 & 0.23 & 9.92 \\

\hline
\end{tabular} \label{tab:fit_paras}
\end{table}

\bibliography{Lohof_et_all_arXiv} 
\end{document}